\documentclass{aa}
\usepackage{graphicx}
\usepackage{txfonts}
\usepackage{natbib}
\usepackage[figuresright]{rotating}
\usepackage{aalongtable,lscape}

\def\cyg{{Cygnus\,OB2}\,}
\def\chandra{{\em Chandra\,}}
\def\einstein{{\em EINSTEIN\,}}
\def\xmm{{\em XMM-Newton\,}}

\def\rosat{{\em ROSAT\,}}

\def\Lx{{L$_{\rm x}$}\,}
\def\Lbol{{L$_{\rm bol}$}\,}

\begin{document}
   \title{X-ray flaring from the young stars in \cyg}
   

   \author{J.F. Albacete Colombo\inst{1,2}
     \and       
     M. Caramazza\inst{1}       
     \and
     E. Flaccomio\inst{1}
     \and
     G. Micela\inst{1}
     \and
     S. Sciortino\inst{1}
   }
   
   \offprints{J.F.A.C. -  email: facundo@astropa.unipa.it}
   
   \institute{INAF - Osservatorio Astronomico di Palermo,
     Piazza del Parlamento 1, I-90134, Palermo, Italy. \and
     Facultad de Ciencias Astron\'omicas y 
     Geof\'isicas de La Plata (FCAGLP), Paseo del Bosque S/N, 
     BFW1900A, La Plata, Argentina}
      
   \date{Received -----; accepted -----}
   
   \abstract{}{We characterize individual and ensemble properties of
X-ray flares from stars in the \cyg and ONC star-forming regions.}{We
analyzed X-ray lightcurves of 1003 \cyg sources observed with Chandra
for 100\,ksec and of 1616 ONC sources detected in the ``Chandra Orion
Ultra-deep Project'' 850\,ksec observation. We employed a binning-free
maximum likelihood method to segment the light-curves into intervals
of constants signal and identified flares on the basis of both the
amplitude and the time-derivative of the source luminosity. We then
derived and compared the flare frequency and energy distribution of
\cyg and ONC sources. The effect of the length of the observation on
these results was investigated by repeating the statistical analysis
on five 100\,ksec-long segments extracted from the ONC data.}{We
detected 147 and 954 flares from the \cyg and ONC sources,
respectively. The flares in \cyg have decay times ranging from
$\lesssim$0.5 to about 10 hours. The flare energy distributions of all
considered flare samples are described at high energies well by a
power law with index $\alpha$=-(2.1$\pm$0.1). At low energies, the
distributions flatten, probably because of detection incompleteness.
We derived average flare frequencies as a function of flare energy.
The flare frequency is seen to depend on the source's intrinsic X-ray
luminosity, but its determination is affected by the length of the
observation. The slope of the high-energy tail of the energy
distribution is, however, affected little. A comparison of \cyg and
ONC sources,  accounting for observational biases, shows that the two
populations, known to have similar X-ray emission levels, have very
similar flare activity.}{Studies of flare activity are only comparable
if performed consistently and taking the observation length into
account. Flaring activity does not vary appreciably between the age of
the ONC ($\sim$1\,Myr) and that of \cyg ($\sim$2\,Myr). The slope of
the distribution of flare energies is consistent with the micro-flare
explanation of the coronal heating.} 

\maketitle


\section{Introduction}  

Pre-main sequence (PMS) stars have high levels of X-ray activity with
non-flaring X-ray luminosities (\Lx) up to 10$^{31}$ ergs/s, about two
magnitude order of above those observed in most main sequence (MS)
stars \citep{2005ApJS..160..582P}. The X-ray activity in the PMS phase
is commonly attributed to a ``scaled up'' solar-like corona formed by
active regions. For non-accreting PMS stars, i.e. weak T-Tauri stars
(WTTSs), the fraction of energy emitted in the X-rays energies with
respect to the bolometric luminosity (\Lbol) is close to the
saturation level,  $\sim$10$^{-3}$, observed for rapidly rotating MS
stars, supporting the idea of a common physical mechanism acting both
in the MS and the PMS phases. A more complex scenario is observed for
PMS stars that are still undergoing mass accretion via magnetically
funneled inflows (classical T-Tauri stars - CTTSs): soft X-rays are
also produced here in accretion shocks \citep[e.g.][]{arg07}, but
coronal activity appears to be somewhat reduced with respect to WTTSs
\citep{2003A&A...402..277F, 2005ApJS..160..582P}. Regardless of their
accretion properties, all PMS stars show the high-amplitude rapid
variability associated with violent magnetic reconnection flares
\citep[e.g.][]{1999ARA&A..37..363F}. 

X-ray variability over a wide range of time scales and scenarios is
common to all magnetically active stars 
\citep[e.g.][]{2003SSRv..108..577F, 2004A&ARv..12...71G,
2006ApJ...649..914S}. On long time scales, this includes rotational
modulation of active regions, their emergence and evolution, and
magnetic cycles \citep[e.g.][]{2003A&A...407L..63M,
2005ApJS..160..450F}. Most of the observed variations, however, have
short time-scales ($\sim$ hours) and can be attributed to the
small-scale flares triggered by magnetic reconnection events. Various
authors have proposed that the overall X-ray emission observed in
magnetically active stars is the result of a large number of
overlapping small flares \citep[e.g.][]{2000ApJ...545.1074D,
2007A&A...471..645C}, in analogy with the micro-flare heating
mechanism proposed for the solar corona
\citep[][]{1991SoPh..133..357H}. The energy distribution of these
events is supposedly described well by a power law, such that the
large majority of flares release only small amounts of energy, so that
we only detect their integrated and time-averaged X-ray emission
\citep[][]{2000ApJ...541..396A,2003ApJ...582..423G}.

In the past, statistical studies of X-ray variability in low-mass
stars were hindered by the poor temporal coverage of \einstein and
\rosat observations, which were often short and very fragmented
\citep[e.g.][]{2003A&A...403..247F}, and also by the limited spectral
coverage of these telescopes. They were essentially limited to soft
energies, while flares are characterized by harder emission. However,
since the launch of the \xmm\, and \chandra\, satellites, these
limitations have been eased. In fact, two major projects relevant to
the study of PMS X-ray activity have been performed recently with the
\chandra and \xmm telescopes, targeting two galactic SFRs:\\  $i$- The
{\em Chandra Orion Ultra-deep Project} (COUP) consisting in an
$\sim$850\,ksec long ($\sim$10 days) \chandra observation of the Orion
Nebula cluster (ONC), performed over a time span of 13 days. With a
total of 1616 detected X-ray sources, the ONC is so far the 
best-studied SFR in X-rays. The COUP observation has been used for a
number of variability studies, among which: flare statistics on 'young
suns' \citep{2005ApJS..160..423W}, physical modeling of intense X-ray
flares \citep{2005ApJS..160..469F}, rotational modulation
\citep{2005ApJS..160..450F}, X-ray variability of hot stars
\cite{2005ApJS..160..557S}, and flaring from very low mass (0.1-0.3
M$_\odot$) stars \citep{2007A&A...471..645C}.\\ $ii$- The \xmm {\it
Extended Survey of Taurus Molecular Clouds} (XEST). In contrast to the
COUP, XEST consists of several different pointings with roughly
uniform continuous exposures of $\sim$ 30\,-\,40 ksec
\citep[][]{2007A&A...468..353G}. An X-ray variability study has 
recently been presented by \cite{2007A&A...468..463S}, focusing on
the  statistics of X-ray flaring sources in the context of the coronal
heating processes.

\cyg is a massive $\sim$2\,Myr old star-forming region whose stellar
population has recently been studied in the X-rays by
\citet{2007A&A...464..211A}. This work, based on a 100\,ksec long {\em
Chandra} observation, focused essentially on the detection of X-ray
point sources, finding 1003 from the analysis of their X-ray spectra
and on their optical and near-IR characterization. Here we present an
X-ray variability study of the young low-mass stars of the \cyg
region, aimed at giving a statistical characterization of variability
and, ultimately, understanding its physical origin. Through a
comparison with other regions, this study will also allow us to
understand whether the X-ray variability properties depend on the
different cluster environment and physical characteristics.
Unfortunately, statistical results obtained for one region often
cannot be directly compared with those obtained for another SFR,
because of the different sensitivity limits and temporal coverages of
the respective X-ray observations. However, using the long COUP
observation for the ONC, here re-analyzed consistently with the \cyg
data, we assess the effect of the observation length on our results
and are able to present a robust comparison of flare variability in
the two regions. 

In \S\,2 we describe the methods used to detect and characterize
variability and, in particular, our operative definition of flares. In
\S\,3 we present the observed flare properties: decay times, energy
distribution and frequency. In \S\,4 we extend our study to the COUP
data, considering both the entire 850 ksec observation and 5 distinct
100 ksec segments. Finally, we discuss our results in \S\,5 and draw
our conclusions .

\section{Variability analysis} 
\label{sect:flares} 

The analysis presented in this paper was performed starting from event
lists of the \cyg and ONC \chandra sources. These were appropriately
extracted in the 0.5-8.0 keV energy range by
\cite{2007A&A...464..211A} and \cite{get05a} for the \cyg and ONC
observations, respectively, in both cases using the ACIS-EXTRACT
package \citep{acisextract}. The analysis of the COUP data was
performed in two ways: $i$- considering the entire 850 ksec COUP
observation, $ii$- by selecting in the observations 5 different 100
ksec segments of continuous observation starting at 0, 180, 400, 650,
and 850 ksec from the beginning of the observation\footnote{Julian
times (UT): 2452648.41, 2452650.58, 2452653.03, 2452655.92, and
2452658.27}. This approach proved useful for understanding the biases
due to differences in the observation lengths. The 850 ksec COUP
observation, for example, permits detection of extremely long duration
($\sim$400 ksec) flares \citep{2005ApJS..160..469F}, while in a 100
ksec observation like the one available for \cyg, we can only detect a
fraction of the flares longer than 30-100 ksec and completely miss
those with exponential decay times longer than $\sim$100 ksec. This
results in a bias against the observation of energetic flares.

We initially searched the time series of each \cyg star for
variability using the one-sided Kolmogorov-Smirnov (KS) test
\citep[][]{1992nrfa.book.....P}. This test compares the distribution
of photon arrival times with what is expected for a constant source
and gives the confidence, P$_{\rm KS}$, with which the hypothesis that
the source is constant can be rejected. Sources with P$_{\rm
KS}$$\geq$99.9\% have been considered as definitively variable, while
those with 99.0\%$\leq$P$_{\rm KS}$$\leq$99.9\% are considered as
probably variable. Sources with P$_{\rm KS}$$<$99.0 are not considered
significantly variable. In as \cyg we found 135 sources out of a total
of 1003 sources with P$_{\rm KS}$$>$99\%, of which 86 have P$_{\rm
KS}$$\ge$99.9\% \citep{2007A&A...464..211A}. In the ONC the figures
for the entire COUP observations (1616 detected sources) are 977 and
886 \citep{get05a}, while for each of the five 100\, ksec segments we
find, on average, 358 and 264 sources with P$_{\rm KS}>$99\% and
P$_{\rm KS}>$99.9\%, respectively. Last two figures clearly show how
the number of sources for which the KS-test detects variability in a
given relatively short (e.g. 100\,ksec) observation is a lower limit
to the total number of variable sources in the region. This is
essentially due to the fact that most of the observed variability is
in the form of flares, i.e. events that are shorter than our
observation and with duty-cycles that are instead considerably longer
\citep{2005ApJS..160..423W} than 100\,ksec. Moreover, small flares
and/or other low-level variability may remain undetected because the
sensitivity of the KS tests critically depends on the source's photon
statistics. 

We intend to specifically study flare variability here. In the next
section we therefore briefly describe statistical tools that, in
addition to indicating variability, provide an objective description
of the time behavior of the X-ray emission. This, in conjunction with
an operative definition based on our {\em a priori} idea of flare as
an impulsive event, allows their efficient and, most importantly,
unbiased detection in our observed lightcurves. 

\subsection{Maximum likelihood block analysis} 

We make use of a maximum likelihood algorithm that, under the
assumption of Poisson noise, splits a light curve into periods of
``constant'' signal, referred to as {\it blocks}. In contrast to more
conventional approaches, this method works directly on the sequence of
photon arrival times and does not require binning. The maximum
likelihood block (MLB) algorithm is described by
\cite{2005ApJS..160..423W}. It has two free parameters: the minimum
number of counts per block (N$_{\rm min}$) and the confidence level
(CL) used in detecting variability. In the analysis of the 850 ksec
COUP data, \cite{2005ApJS..160..423W} use the MLB algorithm with
N$_{\rm min}$=20, because of the high photon statistic of solar-mass
COUP sources. Unfortunately, our \cyg observation is 8.5 times shorter
than that of COUP, and the distance to the region is $\sim$4 times
greater than for the ONC. Most of the \cyg sources have less than 40
photons, compelling us to adopt a different N$_{\rm min}$ for the
analysis.  A careful comparison showed that, for sources with more
than 40 detected photons, variability is detected consistently using
the MLB algorithm with both N$_{\rm min}$=1 (MLB\,1) and N$_{\rm
min}$=20 (MLB\,20). In sources with lower statistics, however,
detection of variability is hampered by MLB\,20 because more than 40
photons are needed to define two different blocks. Thus, in practice,
the MLB\,1 algorithm is more sensitive to small flares and is just as
effective as MLB\,20 for more energetic ones. 

Following \cite{2005ApJS..160..423W}, we classify blocks into three
different groups according to their emission level: $i-$ blocks
compatible with the {\it characteristic count rate} (R$_{\rm char}$),
defined as the most frequent count rate exhibited by the source; $ii-$
{\it elevated} blocks, during which the flux is above the R$_{\rm
char}$ but usually not associated with impulsive events; $iii-$ {\it
very elevated} blocks, marked by significantly elevated flux levels,
often associated with impulsive events. We also make use of a measure
of the time variation in the count rate, i.e. the {\it derivative},
here defined as the ratio of the difference between the count rates of
two successive blocks ($\Delta$R=R$_1$-R$_2$) and the minimum of the
temporal lengths of the two blocks,
$min$($\Delta$t$_1$,$\Delta$t$_2$): 

\begin{equation} 
{dR \over dt} \equiv {|R_1 - R_2| \over min(\Delta t_1,\Delta t_2)} 
\hskip 0.2cm \cdot
\end{equation}

\noindent We then define a flare as a group of non-characteristic
blocks, beginning with a block for which the ratio between the
derivative and characteristic level is above a threshold, here set to
10$^{-4.2}$ s$^{-1}$:  

\begin{equation} 
{dR \over dt}\cdot{1 \over R_{\rm char}} > D_{\rm th} \equiv 10^{-4.2} 
\hskip 0.2cm \cdot 
\end{equation}  

\noindent The duration of the flare, t$_{\rm flr}$, is defined as the
total duration of consecutive {\it elevated} and {\it very elevated}
blocks belonging to the flare. Note that the adopted value of $D_{\rm
th}$ is slightly lower than the one used by both
\cite{2005ApJS..160..423W} and \cite{2007A&A...471..645C}, i.e.
10$^{-4.0}$  s$^{-1}$. The new, slightly lower threshold ensures the
detection of 13 small but evident flares in the \cyg data that would
otherwise remain undetected. Our block classification scheme is
illustrated in Fig.\,1 for a \cyg source with an evident flare. The
top panel shows the light curve, both with fixed binning and in the
MLB representation, while the lower one shows the time derivative at
the interface between blocks.

\begin{figure}[!t]    \label{mlb}     \centering 
\includegraphics[width=8.8cm,angle=0]{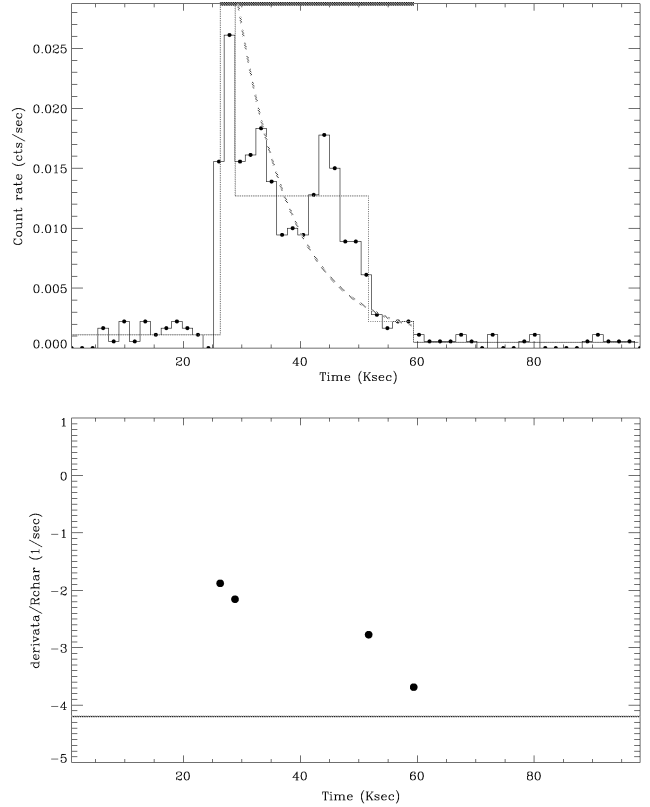}
\caption{{\it Top}: light-curve of a flaring \cyg X-ray source, \#600
in the list of \cite{2007A&A...464..211A}.  The histogram shows the
light-curve binned in 1800 sec bins. The dotted-line  shows the blocks
as computed with the MLB algorithm. The first and last blocks are
compatible with the characteristic level; the second is classified as
``very elevated''; the third and fourth as ``elevated" (see text). The
blocks identified as belonging to the flare are indicated by a thick
solid segment at the top. The dashed line indicates an exponential fit
of decay phase. {\it Bottom}: ratio between the derivative of the
count rate and the characteristic level ${dR \over dt}\cdot{1 \over
R_{\rm char}}$, calculated at the boundaries of each block. The solid
line indicates the threshold (10$^{-4.2}$) used for the detection of
flares. Note: The observed sequential second flare is an example of
the sustained heating and thermodynamics cooling processes competing
simultaneously \citep[e.g.][]{2003SSRv..108..577F}. This results in
longer decay times. MLB compute these cases as single flares.}
\end{figure}

All the results presented throughout this paper are for CL=99.9\%,
N$_{\rm min}$=1, and D$_{\rm th}$=10$^{-4.2}$.

\section{Flare properties of \cyg sources} 
\label{sect:varcyg}  

We applied the MLB algorithm and our operative definition of flare to
the 1003 X-ray sources in the \cyg region. We detected a total of 147
flares occurring in 143 X-ray sources. This means that about 14.7\% of
the sources appear to flare during the 100 ksec \chandra observation.
We found that 74 of the flaring sources are also definitively variable
according to the KS-test (P$_{\rm ks}$$\geq$ 99.9\%), while an other
25 have 99\%$<$P$_{\rm KS}$$<$99.9\%. For the 48 remaining sources,
P$_{\rm KS}$$\leq$99\%. Most of these sources (32/48$\sim$67\%) have
less than 40 photons. Given that the confidence level used for the MLB
algorithm was 99.9\%, this seems to be more efficient than the KS-test
in detecting variability, especially in the low-photon statistical
regime.

\begin{table*}[!ht]
\begin{center}
\label{var_src}
\caption[X-ray properties of flaring sources in the \cyg region. The
complete version is available in the electronic edition of A\&A]{X-ray
properties of flaring sources in the \cyg region. The complete version
is available in the electronic edition of A\&A}
\begin{tabular}{lllllllllllll}
\multicolumn{13}{c}%
{{\bfseries}} \\
\hline \hline \multicolumn{1}{l}{N$_{\rm x}$} &
\multicolumn{1}{l}{Identification} &
\multicolumn{4}{c}{Albacete Colombo et al. (2007)} &
\multicolumn{1}{l}{KS-test} &
\multicolumn{6}{c}{MLB analysis}\\
\cline{3-6}
\cline{8-13}
\multicolumn{1}{l}{\#} &
\multicolumn{1}{l}{2MASS\,J+} &
\multicolumn{1}{l}{Mass} &
\multicolumn{1}{l}{Phot.} &
\multicolumn{1}{l}{E$_{\rm x}$} &
\multicolumn{1}{l}{log(L$_{\rm x}$)} &
\multicolumn{1}{l}{log(P$_{\rm KS}$)}&
\multicolumn{1}{l}{R$_{\rm char}$}&
\multicolumn{1}{l}{T$_{\rm flr}$} &
\multicolumn{1}{l}{log(L$^{\rm peak}_{\rm x}$)}&
\multicolumn{1}{l}{E$_{\rm flr}$} &
\multicolumn{1}{l}{$\tau_{\rm E/L}^{\rm
meas}$} &
\multicolumn{1}{l}{$\tau_{\rm fit}$} \\
\multicolumn{1}{l}{[1]} &
\multicolumn{1}{l}{[2]} &
\multicolumn{1}{l}{[3]} &
\multicolumn{1}{l}{[4]} &
\multicolumn{1}{l}{[5]} &
\multicolumn{1}{l}{[6]} &
\multicolumn{1}{l}{[7]} &
\multicolumn{1}{l}{[8]} &
\multicolumn{1}{l}{[9]} &
\multicolumn{1}{l}{[10]} &
\multicolumn{1}{l}{[11]} &
\multicolumn{1}{l}{[12]} &
\multicolumn{1}{l}{[13]} \\
\hline
   1 &$---------$        &  $--$&   67 &  3.23 & 30.80 & -4.00 &  0.158 &   2.21 &  32.10 & 35.62 &   0.91 &   0.61 \\
   6 &    20322768+4113169 &  1.59 &  216 &  2.61 & 31.32 & -4.00 &  0.958 &   8.18 &  31.98 & 35.99 &   2.87 &   2.64 \\
   8 &    20322913+4114012 &  0.48 &   33 &  2.62 & 30.26 & -0.59 &  0.300 &   0.10 &  31.46 & 34.48 &   0.30 &    $--$ \\
  20 &    20323249+4113131 &  2.27 &   49 &  2.64 & 30.65 & -2.69 &  0.145 &   9.01 &  31.01 & 35.46 &   7.70 &    $--$ \\
  28 &    20323592+4112505 &  0.94 &   43 &  2.04 & 30.51 & -2.14 &  0.245 &   5.03 &  31.15 & 35.20 &   3.06 &    $--$ \\
  33 &    20323661+4122147 &  6.83 &  232 &  1.93 & 31.22 & -4.00 &  0.985 &  12.14 &  31.74 & 36.03 &   5.40 &   4.45 \\
  37 &    20323751+4117121 &  0.18 &   55 &  2.68 & 30.68 & -4.00 &  0.104 &   3.85 &  31.44 & 35.55 &   3.59 &    $--$ \\
  39 &    20323784+4122087 &  1.41 &   91 &  2.20 & 30.81 & -1.56 &  0.463 &   9.87 &  31.15 & 35.58 &   7.43 &    $--$ \\
  43 &    20323810+4112442 &  1.19 &   53 &  2.74 & 30.64 & -4.00 &  0.246 &   3.14 &  31.38 & 35.40 &   2.91 &    $--$ \\
  51 &    20323989+4113488 &  0.58 &   32 &  1.97 & 30.23 & -2.91 &  0.127 &   7.08 &  30.90 & 35.19 &   5.46 &    $--$ \\
  52 &    20324009+4110346 &  7.97 &  156 &  2.69 & 31.28 & -4.00 &  0.533 &   3.72 &  32.04 & 35.93 &   2.14 &   1.62 \\
  54 &    20324032+4122473 &  1.16 &   73 &  2.68 & 30.68 & -1.06 &  0.570 &   1.44 &  31.50 & 35.14 &   1.21 &    $--$ \\
  57 &    20324063+4111113 &  0.54 &   30 &  2.71 & 31.05 & -2.31 &  0.125 &   8.68 &  30.80 & 35.19 &   6.72 &    $--$ \\
  74 &    20324288+4119148 &  1.52 &   89 &  1.86 & 30.98 & -4.00 &  0.238 &   8.08 &  31.44 & 35.72 &   5.25 &   2.82 \\
  81 &    20324342+4111034 &  0.60 &   47 &  2.20 & 30.58 & -4.00 &  0.180 &   5.19 &  31.15 & 35.37 &   4.55 &    $--$ \\
  93 &    20324492+4115203 &  2.43 &   45 &  2.08 & 30.99 & -1.52 &  0.288 &   6.22 &  30.90 & 35.13 &   4.77 &    $--$ \\
 125 &    20324860+4108117 &  1.27 &  121 &  2.23 & 30.95 & -1.33 &  1.165 &   0.13 &  31.68 & 34.77 &   0.34 &    $--$ \\
 135 &    20325011+4116270 &  0.52 &   13 &  2.18 & 26.58 & -1.48 &  0.056 &   3.76 &  31.74 & 34.80 &   0.32 &    $--$ \\
 145 &$---------$        &  $--$&   15 &  2.49 & 30.43 & -2.98 &  0.046 &   4.05 &  30.85 & 34.94 &   3.43 &    $--$ \\
 146 &$---------$        &  $--$&   27 &  3.19 & 30.32 & -0.59 &  0.175 &   2.21 &  31.07 & 34.92 &   1.94 &    $--$ \\
 159 &    20325218+4118160 &  2.83 &   21 &  1.79 & 30.28 & -2.12 &  0.112 &   4.18 &  30.83 & 34.94 &   3.59 &    $--$ \\
 171 &    20325351+4120195 &  0.63 &   11 &  3.85 & 31.99 & -0.89 &  0.016 &   4.72 &  30.65 & 34.89 &   4.83 &    $--$ \\
 172 &    20325377+4115134 &  2.39 &  219 &  2.42 & 31.41 & -4.00 &  0.924 &   9.29 &  31.80 & 36.01 &   4.47 &   3.65 \\
\hline                                                                                                    
\end{tabular}                                                                                              
\end{center}            
                                                                                  
Note: Rows refer to a source flare. Col. 1: source numbers according
to Albacete Colombo et al (2007). Col. 2: name of 2MASS counterpart.
Col. 3-6: source characteristics, i.e. mass of stars, ``phot'' the net
detected photons, ``E$_{\rm x}$'' median photon energy, and ``\Lx''
X-ray luminosity [erg/s]. Col. 7: logarithm of P$_{\rm KS}$. Col. 8
count rate of the characteristic level in units 10$^{-3}$\,[s$^{-1}$].
Col 9: flare duration. Col 10: logarithm of the unabsorbed \Lx at the
flare peak was computed using CF. Col 11: logarithm of the flare
energy (Eq.\,\ref{eq:ene}). Col 12: $\tau_{\rm L/E}$=E$_{\rm
flr}$/L$_{\rm peak}$. Col 13: fitted decay time of the flare.

\end{table*}                                                                                              

In order to estimate the energy of each flare, E$_{\rm flr}$, we
multiply the number of flare photons, $N_{\rm ph-flr}$ (=$\sum
{(R-R_{char})\Delta T}$), by a counts-to-energy conversion factor,
$CF$. This is computed from the total energy released by the source
during the observation, i.e. the mean \Lx times the exposure time,
divided by the number of detected source photons, $N_{\rm ph-src}$:

\begin{equation}
\label{eq:ene}    E_{\rm flr}={N_{\rm ph-flr} \over N_{\rm
ph-src}}\int_{0}^{T_{\rm exp}}L_{\rm x}(t)\,dt= {\overline{L_{\rm
x}}T_{\rm exp} \over N_{\rm ph-src}}N_{\rm ph-flr} = CF\cdot N_{\rm
ph-flr}   \hskip 0.2cm \cdot 
\end{equation}

For \cyg sources $\overline{\rm L_{\rm x}}$ was computed by
\cite{2007A&A...464..211A} using a single count rate for the flux
conversion factor. For ONC, although X-ray luminosities from X-ray
spectral fits for individual sources are available \citep{get05a}, we
derived new \Lx values, for consistency, using the same approach as
was used for \cyg\,. Thus E$_{\rm flr}$ was obtained in the two cases
by multiplying $N_{\rm ph-flr}$ by the relative counts-to-energy CF:
7.52$\times$10$^{32}$ ergs/ph for the ONC and 7.92$\times$10$^{33}$
ergs/ph for \cyg. 

We thus ignored both the source-to-source and the flare-to-flare
variability in X-ray spectra. Given the low photon statistic of the
sources and flares, in particular of those in \cyg, it is indeed not
possible to derive source-specific and flare-specific $CF$s. The
dependence of $CF$ on plasma temperature is, however, expected to be
small.

\subsection{Flare duration and decay times}

Most observed flares appear to decay exponentially. We thus
approximate their light curves as \Lx(t)=L$_{\rm 0}$e$^{-t/\tau}$,
where L$_{\rm 0}$ is the peak luminosity and $\tau$ the exponential
decay time. The parameter $\tau$ is one of most important
observationally accessible since it carries information on the
physical conditions of the flaring plasma and on the geometry of its
confining magnetic field \citep{1991A&A...241..197S,
2004A&A...416..733R}. 

Unfortunately, $\tau$ is not easily measured for most of our flares
because the photon statistic is insufficient for a reliable
exponential fit. In the exponential assumption, the {\em measured}
energy of a flare, E$_{\rm flr}$, which we estimate from the number of
photons detected from the beginning of the rise phase (assumed
instantaneous) to its {\em detected} duration, {\bf T$_{\rm flr}$},
can be written as

\begin{equation}  E_{\rm flr} = \int^{T_{\rm flr}}_0 L_{\rm
0}\,e^{-t/\tau} dt  \,\,\,\,\longrightarrow\,\,\,\,\, {E_{\rm flr}
\over L_{\rm 0}} = \tau(1 - e^{-{\rm T}_{\rm flr}/\tau}) \hskip 0.2cm
\cdot  \label{eq:tauEL} \end{equation} 

\begin{figure}[!t]
\centering  \includegraphics[width=8.8cm,angle=0]{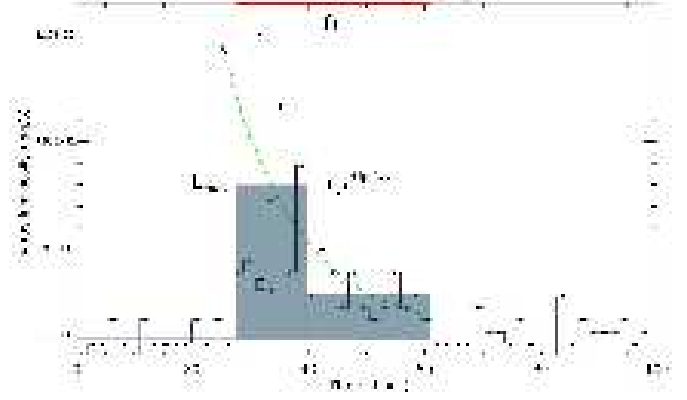}
\caption{We illustrate flare parameters. Observed flare correspond to 
source N$_{\rm x}$\,172. Values of the parameters can be followed in
Table\,\ref{var_src}. Y-axis was converted from the count rate by
using CF=7.92$\times$10$^{33}$ erg/ph. The shadowed area corresponds
to the flare energy. Dashed line refers to the exponential decay fit
of the flare. Note: Label ``tau'' in the equation refers to
exponential decay time $\tau$. For flares characterized by three or
more blocks, we give fitted $\tau_{\rm fit}$ values in
Table\,\ref{var_src}.} \label{clear} 
\end{figure}

\noindent The additive term $e^{-{\rm T}_{\rm flr}/\tau}$ in
Eq.\,\ref{eq:tauEL} could be understood as a correction factor of
$\tau$ in a integration limited by T$_{\rm flr}$. Note that the
quantity ${E_{\rm flr} \over L_{\rm 0}}$ approaches $\tau$ for 
T$_{\rm flr}\gg\tau$. Hereafter, we indicate ${E_{\rm flr} \over
L_{\rm 0}}$ as $\tau_{\rm E/L}$. In Fig\,\ref{clear} we illustrate
computed flare parameters for a typical flare. 

\begin{figure}[!t]
\centering  \includegraphics[width=8.5cm,angle=0]{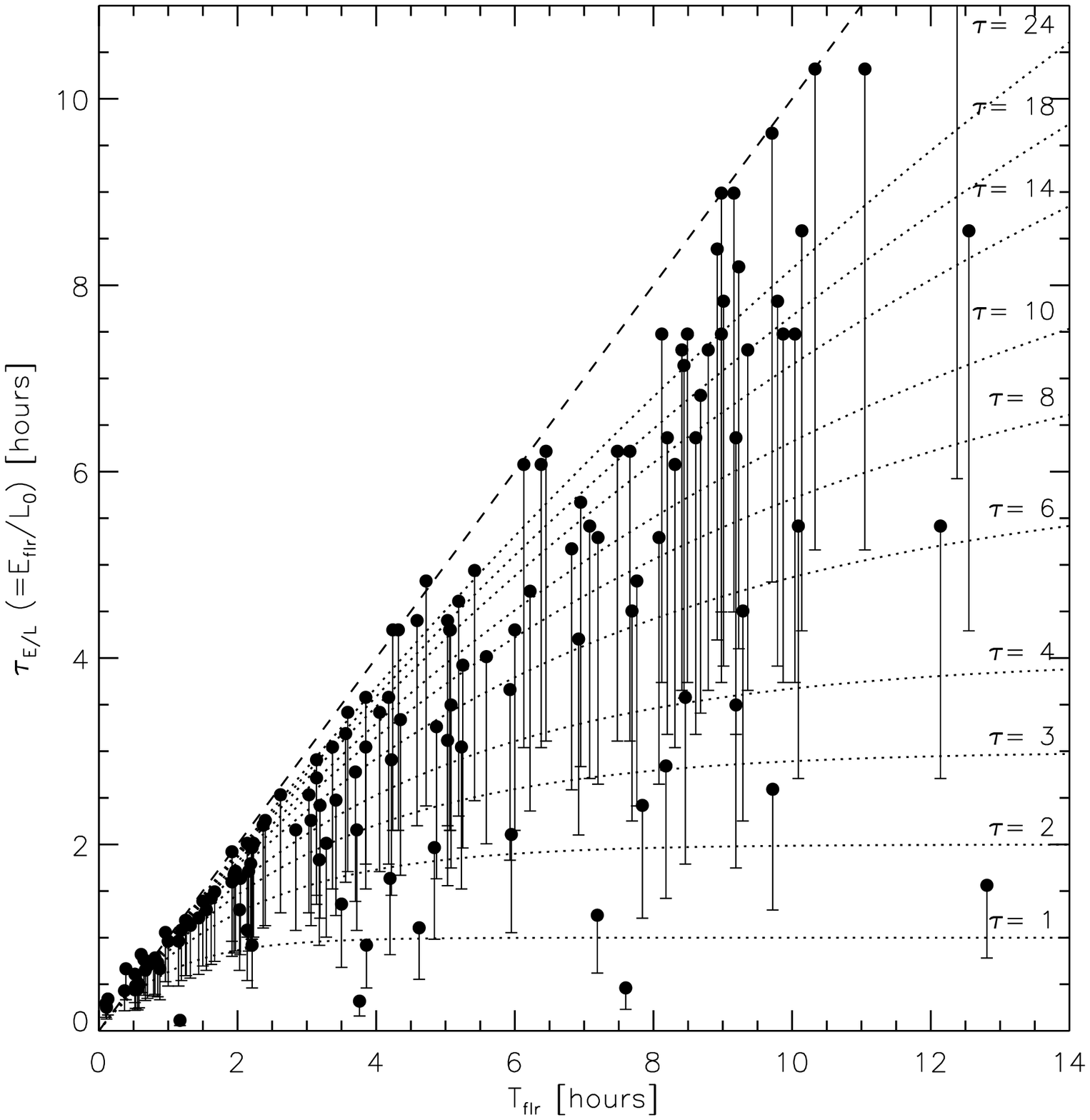}  
\caption{$\tau_{\rm E/L}$($\equiv$E$_{\rm flr}$/L$_{\rm 0}$) vs. 
T$_{\rm flr}$, the measured duration of flares. Filled circles
indicate the observationally derived values, E$_{\rm flr}$/L$_{\rm
max}$, for flares detected on \cyg stars; vertical error-bars the
allowed range of $\tau_{\rm E/L}$ considering the observational biases
on L$_{\rm 0}$: [0.5E$_{\rm flr}$/L$_{\rm max}$,E$_{\rm flr}$/L$_{\rm
max}$]. Curves show the theoretical loci for exponentially decaying
flares (Eq. \ref{eq:tauEL}) with different decay times, $\tau$, as
labeled on the right-hand side (in hours). Note: Dashed line
represents the limit case for which T$_{\rm flr}\gg\tau$ (Eq.
\ref{eq:tauEL}), and thus the exponential correction factor ($e^{-{\rm
T}_{\rm flr}/\tau}$) vanishes (see text).} \label{fig:tau_plot}
\end{figure}

In Fig.\,\ref{fig:tau_plot} we plot $\tau_{\rm E/L}$ vs. the time
duration of flares, t$_{\rm flr}$, both estimated from the MLB
analysis. We note in particular that L$_{\rm 0}$, entering into the
calculation of $\tau_{\rm E/L}$, is approximated here with the maximum
luminosity of the blocked flare light curve, L$_{\rm max}$. For an
impulsive rise followed by an exponential decay, this last quantity is
by definition smaller than the true peak flare luminosity (i.e.
L$_{\rm max}$$<$L$_{\rm 0}$), but typically greater than
$\sim$0.5\,L$_{\rm 0}$. The vertical error bars in
Fig.\,\ref{fig:tau_plot} reflect the resulting uncertainty on
$\tau_{\rm E/L}$: 0.5E$_{\rm flr}$/L$_{\rm max}$$<$$\tau_{\rm
E/L}$$<$E$_{\rm flr}$/L$_{\rm max}$. We also plot the theoretical
loci, calculated from Eq\,\ref{eq:tauEL}, for values of $\tau$ ranging
from 1 to 24 hours. A comparison of these theoretical curves with the
data points indicates that the observed flares span a wide range of
decay times.  

\begin{figure*}[!t]
\centering  \includegraphics[width=18cm,angle=0]{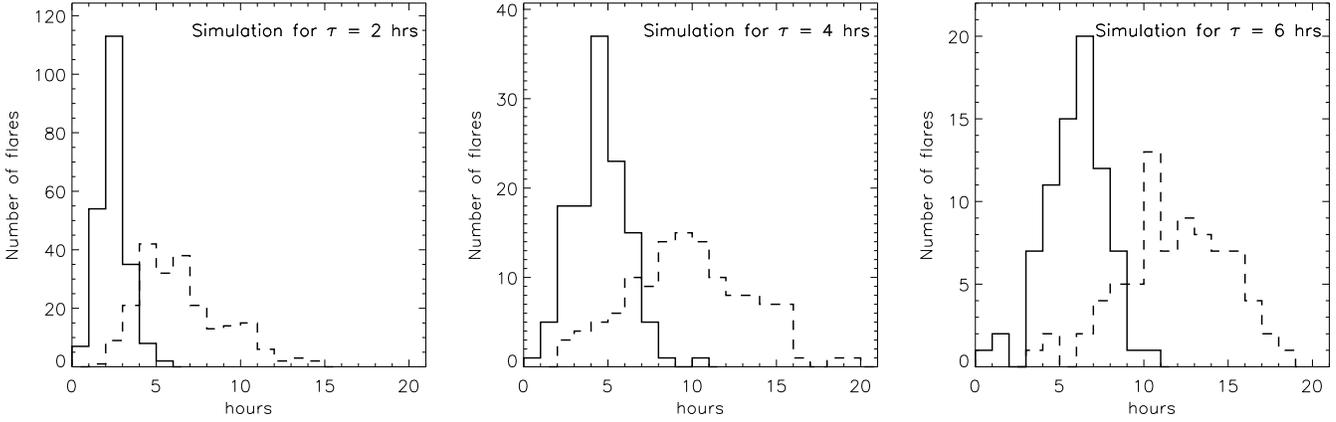}
\caption{Distributions of the detected flare duration t$_{\rm flr}$
(dashed lines) and $\tau_{\rm E/L}^{\rm meas}$ (solid lines) for
simulated light curves constructed with flares having three different
decay times $\tau$: 2,4, and 6 hours. Note that, while the
distributions of $\tau_{E/L}^{\rm meas}$ peak at the simulation input
values, t$_{\rm flr}$ appears broadly distributed towards higher
values. \label{simul_tau}}   
\end{figure*}

We now wish to test the usefulness of $\tau_{\rm E/L}^{\rm
meas}$($\equiv$E$_{\rm flr}$/L$_{\rm max}$) as a measure of the
flare's decay time. For this purpose we have created a large number of
simulated light curves assuming the same flare energy distribution
found for \cyg sources (see \S \ref{stat}) and for a set of $\tau$
ranging from 2 to 16 hours. We then applied the MLB algorithm and our
flare definition to characterize flare properties in the same way as
for \cyg sources. In Fig.\,\ref{simul_tau} we present histograms of
$\tau_{\rm E/L}^{\rm meas}$ and t$_{\rm flr}$ resulting for
simulations with $\tau$ equal to 2, 4, and 6 hours. While t$_{\rm
flr}$ appears broadly distributed, the distribution of $\tau_{\rm
E/L}^{\rm meas}$ for simulated flare values peak at the corresponding
$\tau$ input value. This suggests that $\tau_{\rm E/L}^{\rm meas}$ may
indeed be used as an estimator of $\tau$, with a typical, associated
uncertainty $\sigma$($\tau$)$\sim$0.3$\tau$. Columns 12 and 13 of
Table 1 give the values of $\tau_{\rm E/L}^{\rm meas}$ for all the 147
flares detected on \cyg sources. The median $\tau_{\rm E/L}^{\rm
meas}$ is 2.9 hours and 90\% of the values are in the [0.5,9] hour
range. For 27 flares for which the light curve comprised more than one
block we also computed $\tau$ from exponential fits to the block count
rates. For these 27 sources, the median $\tau_{\rm E/L}^{\rm meas}$
and $\tau_{\rm fit}$ are both 2.6 hours. The distributions of
$\tau_{\rm E/L}^{\rm meas}$ for the whole sample and for the 27 flares
with exponential fits and that of $\tau_{\rm fit}$ for the latter
sample are all statistically indistinguishable from each other using
two-sample Kolmogorov-Smirnov tests (all giving null probabilities
$>$16\%).

\subsection{Flare energy distribution} 
\label{stat}

Studies of the solar corona
\citep[i.e.][]{1984ApJ...283..421L,1998ApJ...501L.213K} have found
solid evidence of small-scale of continuous flaring. The distribution
in energy of these flares was found to follow a power law:
dN/dE\,$\propto$\,E$^{-\alpha}$, where dN is the number of flares
produced in a given time interval with a total energy (thermal and
radiated) in the interval [E, E+dE], and $\alpha$ is the  index of the
power-law distribution
\citep[][]{1974SoPh...39..155D,1991SoPh..133..357H}. 

The total energy released in flares is obtained by integration:

\begin{equation}  E_{\rm tot} = \int^{E_{\rm max}}_{E_{\rm min}}{dN
\over dE}\,E\,dE =  k\int^{E_{\rm max}}_{E_{\rm min}}E^{1-\alpha}\,dE
\approx {k \over \alpha-2}\,E_{\rm min}^{-(\alpha-2)} \hskip 0.2cm
\cdot \end{equation}

\noindent where $E_{\rm min}$ and $E_{\rm max}$ are the minimum and
maximum allowed flare energies, respectively, $k$ is a normalization
factor and in the last passage we assumed $\alpha$$>$2 and $E_{\rm
max}$$\gg$$E_{\rm min}$. For $\alpha$$>$2, the integral diverges if
$E_{\rm  min}$$\rightarrow$0, meaning that the energy released from
micro flares ($\sim$10$^{27}$-10$^{30}$ ergs) and nano flares
($\sim$10$^{24}$-10$^{27}$ ergs) can contribute large amounts of
energy to the total emission and eventually explain the whole coronal
output \citep{1991SoPh..133..357H}. A non-zero value of $E_{\rm min}$
or a flattening of the flare energy distribution at energies is
required to keep the total emitted energy finite. The cumulative
distribution of flare energies also follows a power law: 

\begin{equation}  N(>E_{\rm flr})\propto\int^{\infty}_{\rm E_{\rm
flr}}E^{-\alpha}\,dE\propto{1 \over \beta}{E_{\rm flr}}^{-\beta}  {\rm
\,\,\,\,where\,\,\,\beta=\alpha-1} \hskip 0.2cm \cdot \end{equation}

In the following, we compare the observed cumulative distribution of
flare energies with the above theoretical distribution in order to
constrain the slope, $\alpha$. More precisely, we consider the
observed {\em frequency} of flares per source vs. the minimum flare
energy, i.e. N($>E_{\rm flr}$) divided by the number of sources in our
sample and by the duration of the observation\footnote{We are using
the ergodic principle here, assuming that all the sources in our
sample share the same distribution of flare energies (although they do
not need to share the same rate of flaring). This is necessary in
order to improve the statistics, given the low frequency of flares for
a single source.}. Note however that, even if the simple power law
model is correct, the observed distributions will flatten at low
energies due to an observational bias; i.e., small events are missed
by our flare detection process. Above a certain threshold energy,
E$_{\rm cut}$, however, our list of detected flares will probably be
complete.

\begin{figure}[!ht]
\centering  \includegraphics[width=8.7cm,angle=0]{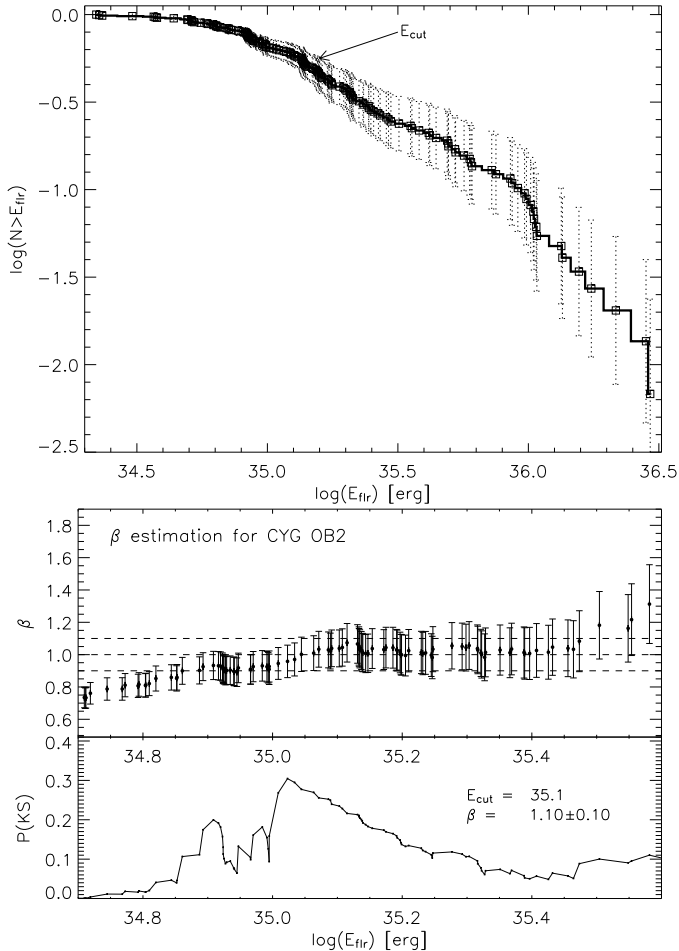} 
\caption{{\it Upper panel}: Frequency of flares (flares per source per
ksec) with energy above E$_{\rm flr}$ vs. E$_{\rm flr}$, for \cyg
sources as derived from our 100 ksec \chandra observation. {\it Middle
panel}: Maximum Likelihood estimate of the power-law index that better
approximate the distribution for flare energies above E$_{\rm flr}$.
Three dashed lines, from top to bottom, ref fer to $\beta$-values 1.1,
1.0, 0.9, respectively. {\it Bottom panel}: probability, according to
the KS test, that a power law with best-fit index as given in the
middle panel, is compatible with the observed distribution for
energies above E$_{\rm cut}$. \label{fig:cumd_cyg}}   
\end{figure}

Other than the sensitivity bias, a simple cumulative distribution of
flare energies can also be biased by the effect of overlapping flares.
More specifically, the ``effective'' time available for the
identification of flares with a given energy is reduced by the
presence of larger ones, usually known as {\it dead-time} correction.
This effect is generally small, given the low frequency of detectable
flares. We correct for this bias, however in the derivation of the
frequency of flares by assuming that the time available for detect 
flares with a given energy is reduced by the sum of the durations of
more energetic flares
\citep[see,][]{2000ApJ...541..396A,2007A&A...468..463S}. Figure
\ref{fig:cumd_cyg} (upper panel) shows, for the \cyg X-ray sources,
the dead-time-corrected cumulative distribution of flare energies, in
units of flares per source per ksec.

We use the maximum likelihood method described by
\citet{1970ApJ...162..405C} to determine E$_{\rm cut}$, i.e. the
minimum energy above which the distribution is compatible with a power
law, and $\beta$, the best-fit slope of the power law. The details of
the statistical analysis can also be followed in
\cite{2007A&A...468..463S}. 

In summary, we have computed, as a function of E$_{\rm flr}$, the
maximum-likelihood slope, $\beta$(E$_{\rm flr}$), of the cumulative
distribution of flares with energies higher than E$_{\rm flr}$.
Alongside $\beta$(E$_{\rm flr}$) we also used the Kolmogorov-Smirnov
test to computed P$_{\rm KS}$(E$_{\rm flr}$), i.e. the probability
that a power law with index $\beta$(E$_{\rm flr}$) is indeed
compatible with the observed distribution. The two functions are shown
in the middle and bottom panels of Fig. \ref{fig:cumd_cyg}. We chose
E$_{\rm cut}$=10$^{35.1}$\,ergs\,s$^{-1}$ with respect to the maximum
of P$_{\rm KS}$ ($\sim$30\%). This leads in an index
$\beta$$\approx$1.1$\pm$0.1. Note that further increasing E$_{\rm
cut}$ means $\beta$(E$_{\rm flr})$ reaches a plateau, confirming the
compatibility of the observed flare energy distribution, above E$_{\rm
cut}$, with a power law.

In spite of our statistical analysis, our estimation of $\beta$ and
E$_{\rm cut}$ from the observed energy flare distribution could still
be biased by the limited exposure time of the observation precluding
the detection of a fraction of the longer flares. This bias could
mainly affect the high-energy tail of the observed flare energy
distribution of \cyg sources. Unfortunately, with our 100 ksec
observation, we are unable to quantify the degree of this
incompleteness. 

In the extraordinarily long COUP observation ($\sim$850 ksec),
however, this bias may be considered negligible. In \S \ref{sect:COUP}
we present a re-analysis of the COUP data aimed at addressing this
issue and at comparing the flare frequency of \cyg stars with those of
the ONC. First, however, we continue to discuss the issue of the
frequency of flares.

\subsection{Flare frequency}  
\label{freq}  

The low/intermediate-mass, X-ray-detected stellar population within
the \cyg region covered by our \chandra observation counts, excluding
26 OB stars, 1003-26=977 likely members \citep{2007A&A...464..211A}.
We exclude the OB stars from our statistics because their X-ray
emission is mostly unrelated to magnetic activity. Our MLB analysis
detected a total of 147 flares during the 100 ksec observation. The
average flare frequency (f$_{\rm flr}$) is thus approximately 1 flare
in $\sim$664 ksec\footnote{\citet{2007A&A...464..211A} indicate that
up to $\sim$80 X-ray sources may be of extragalactic nature. Taking
this contamination into account, the computed flare frequency should
be slightly reduced to 1/610 ksec$^{-1}$.}. Considering only flares
with energies above our completeness limit, E$_{\rm cut}$=10$^{35.1}$
erg, the flare frequency is reduced to f$_{\rm flr}$=1/1320
ksec$^{-1}$.

\begin{figure}[!t]
\centering \includegraphics[width=9.4cm,angle=0]{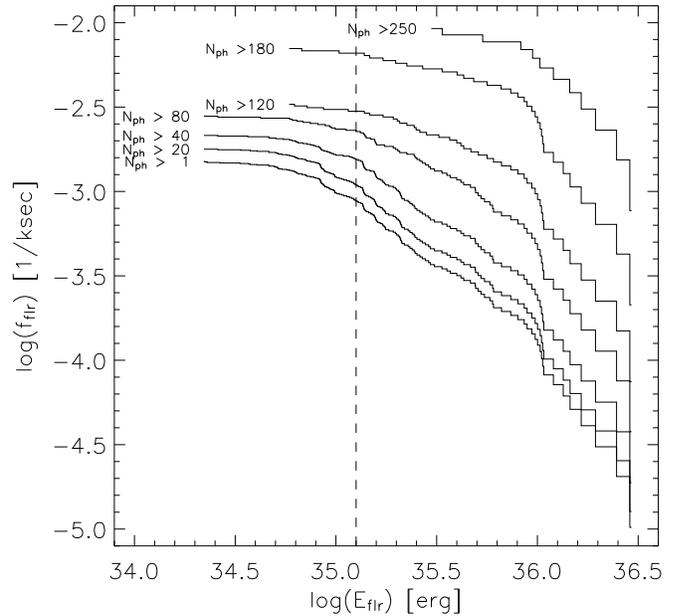}\\
\caption{Flare frequency vs. minimum energy (as in Fig.
\ref{fig:cumd_cyg}) for source sub samples with different minimum
numbers of detected source photons (N$_{\rm ph}$), as indicated beside
each curve. The dashed vertical line marks our flare detection
detection completeness limit determined in \S\,\ref{stat}. Note:  For
sources with a similar N$_{\rm ph}$, differences in the energy of the
flare are due to their respective exponential decay times ($\tau$).}
\label{flr_freq}
\end{figure}

The flare frequency can, however, be expected to be a function of
source photon statistic, because of the combined effects of the
limited sensitivity of our flare detection procedure and of the likely
dependence of the flare intensity distribution on source brightness.
Table \ref{tab_flr} reports the flare frequencies calculated for a sub
sample of stars with different minimum numbers of detected photons.
Figure \ref{flr_freq} also shows the full distributions of flare
frequencies computed for different ranges of the source's photon
statistics. Given the bias against the detection of low-energy flares,
the flare frequencies for different sub samples are more directly
comparable when considering only those flares with energies above our
completeness limit, E$_{\rm flr}$$>$E$_{\rm cut}$. This quantity is
given in the last column of Table \ref{tab_flr}.

\begin{table}[t]
\begin{center}
\caption[Flare frequency for stars with different photon statistics]{Flare frequency 
for stars with different photon statistics \label{tab_flr}}
\begin{tabular}{lllll}
\multicolumn{5}{c}%
{{\bfseries}} \\
\hline \hline 
\multicolumn{1}{l}{Min} &
\multicolumn{1}{c}{N$_{\rm src}$} &
\multicolumn{1}{c}{N$_{\rm flr}$} &
\multicolumn{2}{c}{f$_{\rm flr}$ [ksec$^{-1}$]}\\
\cline{4-5}
\multicolumn{1}{l}{counts} &
\multicolumn{1}{l}{in sample} &
\multicolumn{1}{l}{in sample} &
\multicolumn{1}{l}{total} &
\multicolumn{1}{l}{comp.$^\ast$} \\
\hline
  1 & 977 & 147 & 1/664 & 1/1320\\
 10 & 928 & 143 & 1/644 & 1/1254\\
 20 & 787 & 140 & 1/562 & 1/1063\\
 30 & 630 & 132 & 1/477 & 1/863\\
 40 & 531 & 114 & 1/465 & 1/748\\
 50 & 434 & 102 & 1/425 & 1/629\\
 80 & 265 &  74 & 1/358 & 1/416\\
120 & 134 &  44 & 1/304 & 1/352\\
150 &  78 &  35 & 1/222 & 1/252\\
180 &  47 &  33 & 1/142 & 1/162\\
250 &  13 &  12 & 1/108 & 1/108\\
\hline
\end{tabular}
\end{center}

The 26 X-ray sources identified with OB stars were excluded from the
analysis of flare frequencies.  Note: $\ast$ indicates flare
frequencies (f$_{\rm flr}$)\footnote{f$_{\rm flr}$ were computed as
the inverse of N$_{\rm flr}$/N$_{\rm src}$/100, in units of
ksec$^{-1}$.} with energies above our completeness limit (i.e.
log\,E$_{\rm flr}$$>$35.1 ergs).

\end{table}

Because of the strong dependence of the observed flare frequency on
the sensitivity of the observation and on the intrinsic brightness of
the X-ray sources, a direct comparison with previous determinations
for members of other SFRs is difficult. For example, the flare
frequency (f$_{\rm freq}$) computed for the PMS stars in the Taurus
molecular clouds (TMC) by \citet{2007A&A...468..463S} is
1/200\,ksec$^{-1}$, much higher than our frequency for the whole \cyg
sample. The difference may largely be explained as a sensitivity
effect, given that the TMC is $\sim$11 times closer than \cyg and that
the observation were performed with a more sensitive instrument (\xmm
instead of \chandra). Note that for a ``complete'' sample of flares
(i.e. those with E$_{\rm flr}$$>$10$^{35}$ ergs)
\citet{2007A&A...468..463S} find f$_{\rm flr}$$\sim$1/770 ksec. 
During the COUP observation in the ONC \citet{2005ApJS..160..423W}
observe  41 flares in a sample of 27 solar-mass stars implying f$_{\rm
flr}$$\sim$1/650\,ksec$^{-1}$. This figure reduces to f$_{\rm
flr}$$\sim$1/1150\,ksec$^{-1}$ if we only consider flares with
energies above the completeness limit estimated by
\citet{2007A&A...468..463S}, i.e. 10$^{35.3}$\,ergs.

Obviously, all these rates of flaring are not easily compared because
of the different instruments, exposure times, sensitivity, and the
resulting completeness limits for the observations. The availability
of the COUP data gives us the opportunity to re-analyze this
observation consistently with \cyg data, so as to minimize this 
differences. We give details of this analysis in the next section.

\section{The analysis of the COUP data} 
\label{sect:COUP}

The COUP data set combines six nearly consecutive exposures of the
ONC, spanning 13.2 days (1140 ksec) with a total exposure time of 850
ksec (9.8 days). The observation was taken with the ACIS-I camera
onboard \chandra. A complete description of the COUP data analysis and
source detection procedures can be found in \cite{get05a}. Here we
make use of the 1616 source event files.

\begin{figure*}[!t]
\centering 
\includegraphics[width=17cm,height=6cm,angle=0]{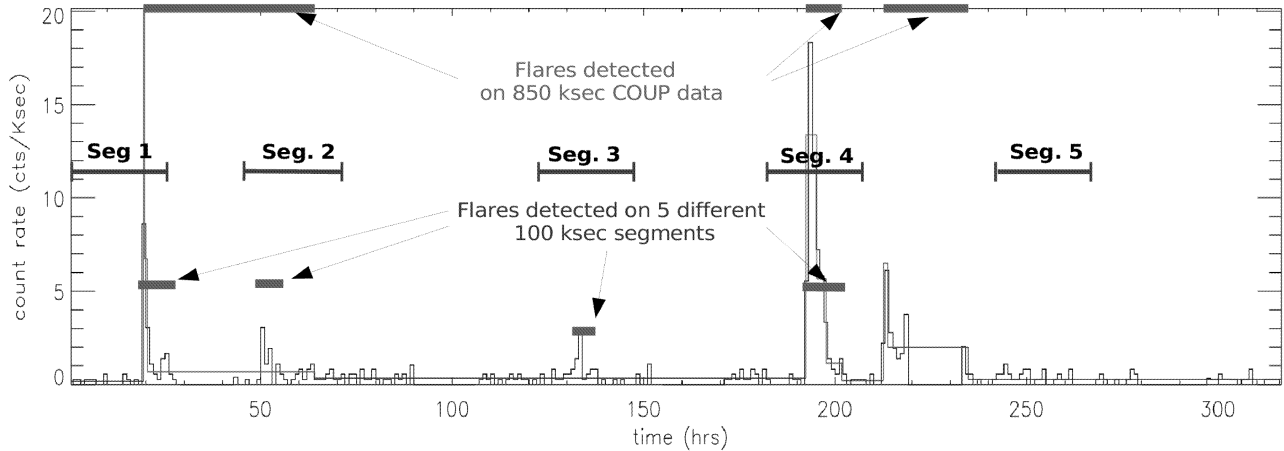} 
\caption{Example of MLB flare detection for COUP source \#1276. The
result of the MLB algorithm applied to the whole observation is
overlaid on the binned light curve. The three thick horizontal
segments on the top indicate flares detected using this
representation.  The four thick segments on the bottom instead
indicate the flares detected by applying the MLB algorithm to the 5
different 100 ksec time intervals, with latter indicated by the
thinner segments labeled ``Seg. 1-5''. Note: time is given in hours
from the beginning of the observation.\label{coup-lc}}  
\end{figure*}

To understand the bias introduced by the limited exposure time of our
100\,ksec \cyg observation, we present the analysis of the entire 850
ksec COUP observation here, as well as that of five different 100 ksec
segments. As anticipated in \S \ref{sect:flares}, the 5$\times$100
ksec segments (Seg. 1,2,3,4, and 5) were defined arbitrarily, but
avoided time gaps in the COUP observation. Figure \ref{coup-lc} shows,
as an example, the light curve of COUP source \#1276. The horizontal
segments labeled as ``Seg. 1-5'' indicate the time intervals for which
we independently performed the flare analysis and we also  indicate
flares detected in the whole observation and in the 5 segments
separately. Because the count rates obey Poison statistics, the
maximum amplitude of fluctuations increases with exposure time. This
implies that the statistical significance of a real signal (e.g. a
flare) is higher when considering a shorter time interval. The MLB
algorithm, applied with the same significance threshold (99.9\% in our
case) to the whole observation and to the shorter segments, can
therefore yield different results. In particular, as exemplified in
Fig.\,\ref{coup-lc}, faint flares may remain undetected in the former
case (seg. 3), and consecutive flares may be detected as a single
event (seg, 1 and 2). 

Inspection of the segmented light curves led to excluding three
sources (COUP \#9, \#828 and \#1462) from the following
considerations, they lie close to the edges of the ACIS-I CCDs:
because of the wobbling of \chandra, the light curves of these sources
show high frequency variations at the wobbling period, leading to the
spurious detection of a large number of flares.

\begin{figure}[t]
\centering  \includegraphics[width=8.7cm,angle=0]{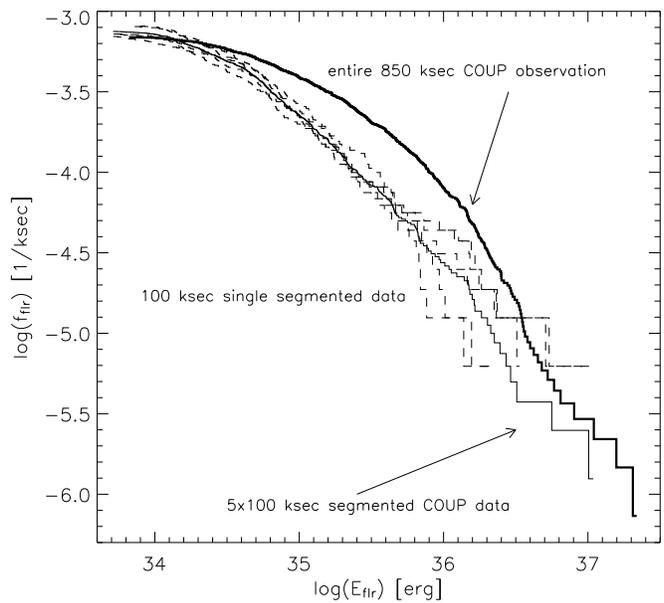} 
\caption{Flare rate vs. minimum flare energy for the ONC X-ray sources
observed by COUP, as determined from flares detected in: $i$- each of
the five selected 100 ksec segments of the COUP data (dashed lines),
$ii$- all the five segments (thin solid line), $iii$- the whole 850
ksec COUP exposure (thick solid line).  Note that, as discussed in the
text, more low-energy flares are detected in the 100\,ksec segments
with respect to the whole observation, but a significant fraction of
the longer/more energetic ones are missed.\label{cumd_coup}} 
\end{figure}

Figure \ref{cumd_coup} shows the rate of flaring vs. minimum flare
energy for the entire 850\,ksec COUP observation, for each of the five
independent 100\,ksec COUP data segments, and for the combined flare
population of the 5 segments. This curve, while maintaining the same
properties with respect to observational biases, such the 5 curves for
the single segments, improves the statistics and allows a more robust
estimation of the mean flare frequency and of slope $\alpha$.  

Results of the MLB analysis for the COUP data are presented in
Table\,\ref{coup-flr}. We report the following for each of the above
data selections: N$_{\rm src}$, the total number of sources that show
flare activity (row 1); N$_{\rm flr}$, the total number of detected
flares (row 2); 1/f$_{\rm flr}$, the inverse of the single-source
flare frequency, i.e. the number of flares divided by the total
observing time in ksec (row 3); E$_{\rm cut}$, the flare energy
threshold above which flare detection is considered statistically
complete (row 4); N$^{\rm comp}_{\rm flr}$, the number of flares with
energies above E$_{\rm cut}$ (row 5); 1/f$^{\rm comp}_{\rm flr}$, the
inverse of the flare frequency for flares with energies above E$_{\rm
cut}$ (row 6); $\alpha$, the slope of the power law that characterizes
the flare energy distribution above E$_{\rm cut}$ (row 7);
$\sigma(\alpha)$,  the 1$\sigma$ uncertainty on $\alpha$ (row 8).
Columns 2 to 6 give results for the analysis of the 5$\times$100 ksec
segments, column 7 the results for the summed 5$\times$100 ksec
segments, and column 8 the results for the analysis of the entire 850
ksec COUP observation.

{\small
\begin{table}[t]

\caption[Statistics of X-ray flares in the ONC]
{Statistics of X-ray flares in the ONC\label{coup-flr}}
\begin{tabular}{lrrrrrrr}
\multicolumn{8}{l}%
{{\bfseries}} \\
\hline \hline 
\multicolumn{1}{l}{COUP} &
\multicolumn{6}{c}{5 $\times$ 100 ksec segments}&
\multicolumn{1}{l}{850}  \\
\cline{2-6}
\multicolumn{1}{l}{Results} &
\multicolumn{1}{l}{Seg1} &
\multicolumn{1}{l}{Seg2} &
\multicolumn{1}{l}{Seg3} &
\multicolumn{1}{l}{Seg4} &
\multicolumn{1}{l}{Seg5} &
\multicolumn{1}{l}{Total} &
\multicolumn{1}{l}{ksec}\\
\hline
N$_{\rm src}$       &  125 &   111&  110 &  111&   127 &   584 &  640\\
N$_{\rm flr}$           &  128 &   112&  112 &  115&   130 &   601 &  954\\
1/f$_{\rm flr}$         &  1253&  1432& 1382 & 1395&  1233 &  1334 & 1429\\   
logE$_{\rm cut}$        & 34.6 &  34.6& 34.5 & 34.6&  34.5 &  34.7 & 35.6\\
N$^{\rm comp}_{\rm flr}$&  75 &     63&  79  &  62 &    89 &   313 &  246\\
1/f$^{\rm comp}_{\rm flr}$& 2138& 2546& 2030 & 2587&  1802 &  2562 & 5542\\ 
$\alpha$                &  1.9 &   1.9& 1.85 &  1.9&   1.9 &   1.9 & 2.05\\
$\sigma(\alpha)$        &  0.1 &   0.1&  0.1 &  0.1&   0.1 &   0.1 & 0.15\\
\hline
\end{tabular}

\medskip

Notes: 1/f$_{\rm flr}$ is in ksec, E$_{\rm cut}$ in erg. COUP sources
\#9, \#828, and \#1462 were excluded from the analysis (see text).

\end{table}}

For the 1616 COUP sources observed for the whole 850 ksec exposure, we
found that detection of flares is complete for events with log(E$_{\rm
flr}$)$>$35.6, and the cumulative distribution follows a power law
with index $\alpha$$\sim$2.05$\pm$0.15, in agreement with results
obtained by \cite{2007A&A...468..463S} for solar-mass stars (i.e.
log(E$_{\rm flr}$)$>$35.3 and $\alpha$$\sim$1.9$\pm$0.2).  For the 100
ksec segments, we obtained $\alpha$=1.9$\pm$0.1, which is within
$\sim$1$\sigma$ of the result obtained for the whole observation. Our
analysis thus indicates that the slope of the flare energy
distribution obtained from a 100 ksec observation is statistically
consistent with what is obtained from a much longer (850 ksec)
observation. The flare frequencies obtained from the shorter exposures
are, however, significantly reduced. An inspection of the results of
the flare detection processes reveals that many flares are missed in
the short exposures either because the segment does not include the
phases characterized by a high time derivative (the rise phase and the
beginning of the decay) or, in a considerable number of cases, because
the characteristic level is not observed and/or correctly determined.
In some other cases, moreover, the flare is detected, but a large part
of the flare falls outside the segment, thus underestimating its
energy.

\subsection{Comparing the ONC with the \cyg region}

\begin{figure}[!t]
\centering  \includegraphics[width=8.7cm,angle=0]{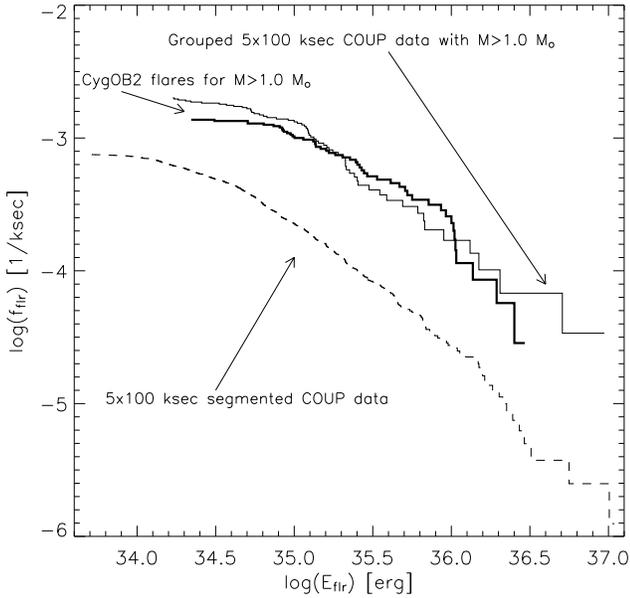} 
\caption{Rate of flaring vs. minimum flare energy for: $i$- the \cyg
sources with mass $\ge$\,1\,M$_\odot$ observed during our 100\,ksec
observation (thick solid line), $ii$- all the COUP sources observed
during the five 100\,ksec segments of the COUP observation (dashed
line), $iii$- COUP sources with mass $\ge$\,1\,M$_\odot$ in the five
100\,ksec segments (thin solid line).\label{onc_cyg_mass}}
\end{figure}

Due to the small difference in age between the ONC (1\,Myr) and \cyg
(2\,Myr), we expect the X-ray properties of stars in the two regions
to be similar. Albacete Colombo et al. (2007) indeed found similar
average X-ray emission levels. For the comparison of flare properties
to be meaningful, we must consider results obtained from observations
of the same temporal length. For the ONC we therefore consider the
results from the the analysis of the five 100\,ksec segments.
Moreover, given the dependence of the X-ray luminosity on stellar mass
and the relation between source luminosity and flare frequency (\S
\ref{freq}), a meaningful comparison between the two regions should
only consider sources in a similar range of mass. We therefore
computed the flare energy distribution for the 92 ONC stars in the
catalog of \citet{get05a} with masses $\ge$ 1M$_\odot$, i.e. roughly
the lower mass limit of X-ray detected \cyg stars
\citep{2007A&A...464..211A}. Figure \ref{onc_cyg_mass} compares this
distribution with that of the \cyg sources with mass $\ge$1M$_\odot$,
both normalized to yield the average single-source rate of flaring.

We analyzed the flare energy distribution for ONC sources with
M\,$\ge$\,1M$_\odot$ using the same statistical procedure as described
in \S\,3. We find that, for E$_{\rm flr}$$\geq$10$^{35.2}$ ergs, the
distributions for \cyg and ONC sources with masses higher than
1M$_\odot$ (see {\it thick} and {\it thin} lines in
Fig.\,\ref{onc_cyg_mass}, respectively) are compatible with a power
law with index $\beta$=1.1$\pm$0.1 (i.e. $\alpha$$\sim$2.1$\pm$0.1).
We conclude that the flare statistics of the \cyg and ONC stars are
very similar.

\section{Summary and conclusions}

We conducted a systematic analysis of flare variability in the soft
X-ray band of the young stars in the \cyg star-forming region. For
this purpose we analyzed the lightcurves of 1003 X-ray sources
detected in a 100\,ksec \chandra observation. For comparison we also 
used the same method to analysis of lightcurves of the 1616 X-ray
sources detected in the ONC with the 850\,ksec observation of the
\chandra Orion Ultra-deep Project. To compare the results for the two
regions, avoiding biases due to the different exposure times, we also
analyzed, independently, five 100\,ksec long segments of the COUP
observation.

Flares were detected using the MLB algorithm. A total of 147 flares
were detected in the lightcurves of 143 \cyg sources. In the ONC, 954
flares were detected on 640 sources in the analysis of the whole
850\,ksec observation, while 601 flares were detected on 584 source in
the analysis of the five 100\,ksec segments.

For each flare we estimated the emitted energy, E$_{\rm flr}$, and the
peak luminosity, L$_{\rm peak}$. Backed by the analysis of extensive
sets of simulated lightcurves, we suggest that the ratio E$_{\rm
flr}$/L$_{\rm peak}$ can be considered a reasonable estimate of the
flare decay-time, $\tau$. This is particularly useful for weak flares,
for which a fit to the light curve is not possible. The 147 flares in
our \cyg sample have a wide range of decay times from $\sim$0.5 hours
to $\sim$10 hours, with a median of 2.9 hours. For the 27 flares that
are bright enough, we fitted the decay phase with an exponential
finding consistent results: a median $\tau$ of 2.6 hours, and a
distribution compatible with that of the whole sample. 

We find that \cyg and ONC flare energy distributions display
high-energy tails described by a power law
(dN/dE$\propto$E$^{-\alpha}$). For energies below a given E$_{\rm
cut}$, the distributions flatten as a result of the incomplete
detection of faint flares. For \cyg sources, we obtain E$_{\rm
cut}$=10$^{35.1}$ ergs and $\alpha$=2.1$\pm$0.1. This slope agrees
with the range of values found in previous studies of solar and
stellar flares \citep{2003ApJ...582..423G,2007A&A...468..463S} and
gives support to the micro-flare hypothesis for the explanation of the
observed X-ray emission and for the heating of corona
\citep{1991SoPh..133..357H}.

The average frequency of flares detected on any given \cyg source
during our 100\,ksec observation is $\sim$1/664\,ksec$^{-1}$. It
reduces to $\sim$1/1320\,ksec$^{-1}$ when considering only flares with
E$_{\rm flr}$$>$E$_{\rm cut}$=10$^{35.1}$ ergs, i.e. those for which
detection is likely to be complete. It is, moreover, important to
stress that our results, even for flares with energies higher than the
``completeness limits'', critically depend on the flare detection
method and on the duration of the X-ray observation. 

We investigated this point using the 850 ksec COUP data set, as well
as  for five distinct 100 ksec segments of the same observation. The
frequencies as a function of minimum flare energy derived in the two
cases are significantly different: the short exposures indeed hinder
the detection of long and (usually) energetic flares, often because it
is not possible to correctly determine the ``characteristic level'',
which is instead clearly observed in the longer exposure. The
``completeness limits'', E$_{\rm cut}$, above which the flare energies
appear to follow power law distributions, are also different for the
two cases: $10^{35.6}$ and $10^{34.6}$ egrs for the 850\,ksec and the
100\,ksec observations, respectively. The slopes of the power laws,
$\alpha$, are however notably similar, respectively 2.05$\pm$0.15 and
1.9$\pm$0.1, which are compatible with each other and with the slope
of the \cyg distribution within $\sim$1$\sigma$. 

For the ONC stars we considered the analysis of the 100\,ksec
segments. To compare stars with similar intrinsic X-ray luminosities,
we restricted the sample to stars with masses higher than
1\,M$_\odot$, i.e. roughly the completeness limit of our \cyg
observation. We find that the stars in the \cyg and ONC regions have
indistinguishable X-ray flare properties.

Finally, we confirm that comparison of flare frequencies is only
allowed if observational limitations and data analysis are performed
in a single and homogeneous way. Contrary to this result is just the
determination of the slope of the power-law distribution, which is not
critically influenced by the length of the observation. 

\begin{acknowledgements}

J.F.A.C acknowledges support by the Marie Curie Fellowship Contract
No. MTKD-CT-2004-002769 of the project ``{\it The Influence of Stellar
High Energy Radiation on Planetary Atmospheres}'' and the host
institution INAF - Osservatorio Astronomico di Palermo. J.F.A.C. also
acknowledges support by the Consejo Nacional de Investigaciones 
Cient\'ificas y Tecnol\'ogicas (CONICET) - Argentina. E.\,F., G.\,M.,
and S.\,S. acknowledge financial support from the Ministero
dell'Universita' e della Ricerca and ASI/INAF Contract I/023/05/0.

\end{acknowledgements}

\bibliographystyle{astron}
\bibliography{8064text}

\begin{thebibliography}{}

\bibitem[\protect\astroncite{{Albacete Colombo}
  et~al.}{2007}]{2007A&A...464..211A}
{Albacete Colombo}, J.~F., {Flaccomio}, E., {Micela}, G., {Sciortino}, S., \&
  {Damiani}, F.: 2007,
\newblock {\em \aap} {\bf 464}, 211

\bibitem[\protect\astroncite{{Argiroffi} et~al.}{2007}]{arg07}
{Argiroffi}, C., {Maggio}, A., \& {Peres}, G.: 2007,
\newblock {\em \aap} {\bf 465}, L5

\bibitem[\protect\astroncite{{Audard} et~al.}{2000}]{2000ApJ...541..396A}
{Audard}, M., {G{\"u}del}, M., {Drake}, J.~J., \& {Kashyap}, V.~L.: 2000,
\newblock {\em \apj} {\bf 541}, 396

\bibitem[\protect\astroncite{{Broos} et~al.}{2002}]{acisextract}
{Broos}, P., {Townsley}, L., {Getman}, K., \& {Bauer}, F.: 2002,
\newblock {\em {ACIS Extract, An ACIS Point Source Extraction Package}},
\newblock http://www.astro.psu.edu/xray/docs/TARA/

\bibitem[\protect\astroncite{{Caramazza} et~al.}{2007}]{2007A&A...471..645C}
{Caramazza}, M., {Flaccomio}, E., {Micela}, G., {Reale}, F., {Wolk}, S.~J., \&
  {Feigelson}, E.~D.: 2007,
\newblock {\em \aap} {\bf 471}, 645

\bibitem[\protect\astroncite{{Crawford} et~al.}{1970}]{1970ApJ...162..405C}
{Crawford}, D.~F., {Jauncey}, D.~L., \& {Murdoch}, H.~S.: 1970,
\newblock {\em \apj} {\bf 162}, 405

\bibitem[\protect\astroncite{{Datlowe} et~al.}{1974}]{1974SoPh...39..155D}
{Datlowe}, D.~W., {Elcan}, M.~J., \& {Hudson}, H.~S.: 1974,
\newblock {\em \solphys} {\bf 39}, 155

\bibitem[\protect\astroncite{{Drake} et~al.}{2000}]{2000ApJ...545.1074D}
{Drake}, J.~J., {Peres}, G., {Orlando}, S., {Laming}, J.~M., \& {Maggio}, A.:
  2000,
\newblock {\em \apj} {\bf 545}, 1074

\bibitem[\protect\astroncite{{Favata} et~al.}{2005}]{2005ApJS..160..469F}
{Favata}, F., {Flaccomio}, E., {Reale}, F., {Micela}, G., {Sciortino}, S.,
  {Shang}, H., {Stassun}, K.~G., \& {Feigelson}, E.~D.: 2005,
\newblock {\em \apjs} {\bf 160}, 469

\bibitem[\protect\astroncite{{Favata} \& {Micela}}{2003}]{2003SSRv..108..577F}
{Favata}, F. \& {Micela}, G.: 2003,
\newblock {\em Space Science Reviews} {\bf 108}, 577

\bibitem[\protect\astroncite{{Feigelson} \&
  {Montmerle}}{1999}]{1999ARA&A..37..363F}
{Feigelson}, E.~D. \& {Montmerle}, T.: 1999,
\newblock {\em \araa} {\bf 37}, 363

\bibitem[\protect\astroncite{{Flaccomio} et~al.}{2003}]{2003A&A...402..277F}
{Flaccomio}, E., {Micela}, G., \& {Sciortino}, S.: 2003,
\newblock {\em \aap} {\bf 402}, 277

\bibitem[\protect\astroncite{{Flaccomio} et~al.}{2005}]{2005ApJS..160..450F}
{Flaccomio}, E., {Micela}, G., {Sciortino}, S., {Feigelson}, E.~D., {Herbst},
  W., {Favata}, F., {Harnden}, Jr., F.~R., \& {Vrtilek}, S.~D.: 2005,
\newblock {\em \apjs} {\bf 160}, 450

\bibitem[\protect\astroncite{{Fuhrmeister} \&
  {Schmitt}}{2003}]{2003A&A...403..247F}
{Fuhrmeister}, B. \& {Schmitt}, J.~H.~M.~M.: 2003,
\newblock {\em \aap} {\bf 403}, 247

\bibitem[\protect\astroncite{{Getman} et~al.}{2005}]{get05a}
{Getman}, K.~V., {Flaccomio}, E., {Broos}, P.~S., {Grosso}, N., {Tsujimoto},
  M., {Townsley}, L., {Garmire}, G.~P., {Kastner}, J., {Li}, J., {Harnden},
  Jr., F.~R., {Wolk}, S., {Murray}, S.~S., {Lada}, C.~J., {Muench}, A.~A.,
  {McCaughrean}, M.~J., {Meeus}, G., {Damiani}, F., {Micela}, G., {Sciortino},
  S., {Bally}, J., {Hillenbrand}, L.~A., {Herbst}, W., {Preibisch}, T., \&
  {Feigelson}, E.~D.: 2005,
\newblock {\em \apjs} {\bf 160}, 319

\bibitem[\protect\astroncite{{G{\"u}del}}{2004}]{2004A&ARv..12...71G}
{G{\"u}del}, M.: 2004,
\newblock {\em \aapr} {\bf 12}, 71

\bibitem[\protect\astroncite{{G{\"u}del} et~al.}{2003}]{2003ApJ...582..423G}
{G{\"u}del}, M., {Audard}, M., {Kashyap}, V.~L., {Drake}, J.~J., \& {Guinan},
  E.~F.: 2003,
\newblock {\em \apj} {\bf 582}, 423

\bibitem[\protect\astroncite{{G\"udel} et~al.}{2007}]{2007A&A...468..353G}
{G\"udel}, M., {Briggs}, K.~R., {Arzner}, K., {Audard}, M., {Bouvier}, J.,
  {Feigelson}, E.~D., {Franciosini}, E., {Glauser}, A., {Grosso}, N., {Micela},
  G., {Monin}, J.-L., {Montmerle}, T., {Padgett}, D.~L., {Palla}, F.,
  {Pillitteri}, I., {Rebull}, L., {Scelsi}, L., {Silva}, L., {Skinner}, S.~L.,
  {Stelzer}, B., \& Telleschi, A.: 2007,
\newblock {\em \aap} {\bf 468}, 453

\bibitem[\protect\astroncite{{Hudson}}{1991}]{1991SoPh..133..357H}
{Hudson}, H.~S.: 1991,
\newblock {\em \solphys} {\bf 133}, 357

\bibitem[\protect\astroncite{{Krucker} \& {Benz}}{1998}]{1998ApJ...501L.213K}
{Krucker}, S. \& {Benz}, A.~O.: 1998,
\newblock {\em \apjl} {\bf 501}, L213+

\bibitem[\protect\astroncite{{Lin} et~al.}{1984}]{1984ApJ...283..421L}
{Lin}, R.~P., {Schwartz}, R.~A., {Kane}, S.~R., {Pelling}, R.~M., \& {Hurley},
  K.~C.: 1984,
\newblock {\em \apj} {\bf 283}, 421

\bibitem[\protect\astroncite{{Marino} et~al.}{2003}]{2003A&A...407L..63M}
{Marino}, A., {Micela}, G., {Peres}, G., \& {Sciortino}, S.: 2003,
\newblock {\em \aap} {\bf 407}, L63

\bibitem[\protect\astroncite{{Preibisch} et~al.}{2005}]{2005ApJS..160..582P}
{Preibisch}, T., {McCaughrean}, M.~J., {Grosso}, N., {Feigelson}, E.~D.,
  {Flaccomio}, E., {Getman}, K., {Hillenbrand}, L.~A., {Meeus}, G., {Micela},
  G., {Sciortino}, S., \& {Stelzer}, B.: 2005,
\newblock {\em \apjs} {\bf 160}, 582

\bibitem[\protect\astroncite{{Press} et~al.}{1992}]{1992nrfa.book.....P}
{Press}, W.~H., {Teukolsky}, S.~A., {Vetterling}, W.~T., \& {Flannery}, B.~P.:
  1992,
\newblock {\em {Numerical recipes in FORTRAN. The art of scientific
  computing}},
\newblock Cambridge: University Press, |c1992, 2nd ed.

\bibitem[\protect\astroncite{{Reale} et~al.}{2004}]{2004A&A...416..733R}
{Reale}, F., {G{\"u}del}, M., {Peres}, G., \& {Audard}, M.: 2004,
\newblock {\em \aap} {\bf 416}, 733

\bibitem[\protect\astroncite{{Serio} et~al.}{1991}]{1991A&A...241..197S}
{Serio}, S., {Reale}, F., {Jakimiec}, J., {Sylwester}, B., \& {Sylwester}, J.:
  1991,
\newblock {\em \aap} {\bf 241}, 197

\bibitem[\protect\astroncite{{Stassun} et~al.}{2006}]{2006ApJ...649..914S}
{Stassun}, K.~G., {van den Berg}, M., {Feigelson}, E., \& {Flaccomio}, E.:
  2006,
\newblock {\em \apj} {\bf 649}, 914

\bibitem[\protect\astroncite{{Stelzer} et~al.}{2007}]{2007A&A...468..463S}
{Stelzer}, B., {Flaccomio}, E., {Briggs}, K., {Micela}, G., {Scelsi}, L.,
  {Audard}, M., {Pillitteri}, I., \& {G\"udel}, M.: 2007,
\newblock {\em \aap} {\bf 468}, 463

\bibitem[\protect\astroncite{{Stelzer} et~al.}{2005}]{2005ApJS..160..557S}
{Stelzer}, B., {Flaccomio}, E., {Montmerle}, T., {Micela}, G., {Sciortino}, S.,
  {Favata}, F., {Preibisch}, T., \& {Feigelson}, E.~D.: 2005,
\newblock {\em \apjs} {\bf 160}, 557

\bibitem[\protect\astroncite{{Wolk} et~al.}{2005}]{2005ApJS..160..423W}
{Wolk}, S.~J., {Harnden}, Jr., F.~R., {Flaccomio}, E., {Micela}, G., {Favata},
  F., {Shang}, H., \& {Feigelson}, E.~D.: 2005,
\newblock {\em \apjs} {\bf 160}, 423

\end{thebibliography}

\end{document}